\documentclass[aps,twocolumn,secnumarabic,nobalancelastpage,amsmath,amssymb,nofootinbib]{revtex4-1}
\usepackage{graphicx}      
\usepackage{epsf}
\usepackage{epstopdf} 
\usepackage{longtable}     
\usepackage{url}           
\usepackage{bm}            
\usepackage{float}
\usepackage{cleveref}

\Crefformat{equation}{Eq.(#1)}
\Crefformat{figure}{Fig. #1}
\Crefformat{table}{Tab. #1}
\Crefformat{section}{Sec. #1}
\Crefformat{subsection}{Sec. #1}
\Crefformat{subsubsec}{Sec. #1}
\Crefrangeformat{equation}{Eqs. #3(#1)#4-#5(#2)#6}
\Crefrangeformat{figure}{Figs. #3#1#4-#5#2#6}

\begin{document}
\title{Ordering effects in 2D hexagonal systems of binary and ternary BCN alloys}
\author         {Agnieszka Jamroz and Jacek A. Majewski}
\email          {ajamroz@fuw.edu.pl}
\affiliation    {Warsaw University, Faculty of Physics}
\date{\today}

\begin{abstract}
We present theoretical study of ordering phenomena in binary $C_{1-x}B_{x}$, $C_{1-x}N_{x}$ and ternary $B_{x}C_{1-x-y}N_{y}$ alloys forming two-dimensional, graphene-like systems. For calculating energy of big systems (20 000 atoms in the supercell with periodic boundary conditions assumed) empirical Tersoff potential was employed. In order to find equilibrium distribution of different species corresponding to minima of the energy, we use Monte Carlo approach in Metropolis regime. We take into consideration wide range of concentrations (1-50\%) and temperatures (70-1500 K), to provide more complete picture. For quantitative description of order, we determine Warren-Cowley Short Range Order (SRO) parameters for the first coordination shell. This procedure allows us to determine energetically favorable compositions of all alloys, and characterize resulting types of order for both binary and ternary systems.
\end{abstract}

\maketitle


\section{Introduction}
Graphene and graphene-based materials have been vigorously investigated during recent years. As a single layer of graphite, graphene is composed of carbon atoms stacked in two-dimensional honeycomb structure, which leads to a number of superb properties, e.g. extreme electron mobility, thermal conductivity and tensile strength, that are desirable for application in electronics, optoelectronics, solar cell systems, etc. However, there is one difficulty to be overcome - semimetalic character of graphene caused by its zero band-gap. Hence, for a while, scientists have been working on finding an effective way of tailoring the gap without big deterioration of material's properties. One of possible solutions could be chemical doping, i.e. introducing atoms of different species into the carbon system. \\
In this study, we focus on substitutional doping with boron and nitrogen - species that are the closest to carbon in periodic table. Various methods to enrich graphene layer with nitrogen atoms are known, among them electrothermal reactions with ammonia \cite{Wang} or low-energy implantation of nitrogen in graphene on a substrate \cite{L-ENI, Willke}. There is also a report on B- and N- implantation of free standing graphene \cite{Banglert}. But there are only a few papers on producing B-doped graphene layer \cite{LiuX, ZhaoB}  (see also Ref. \cite{Humberto} for more details). A lot of experimental research was done in the area of boron-carbon-nitride hexagonal monolayer alloys e.g. \cite{LijieCi, Sutter}. On the other side, wide spectrum of computational results in the topic of substitutional doping of graphene with B and N can be reported, e.g. \cite{Ahlgren,Yuge,Lara}. However, vast part of them are first-principles calculations, that are based on accurate but expensive electron-density calculations, thus they are restricted to few chosen atomic configurations with maximum several hundreds of atoms in the supercell. For the topic of ordering phenomena, it is important to investigate more general cases, that are not influenced by artificial periodic boundary conditions or edges. It is known and expected \cite{Panchcarla} that arrangement of substituting atoms may influence electronic and elastic properties of graphene. Thus it is important to track down types of distribution that are energetically favorable and stable, ergo may exist in reality. Knowledge of ordering character of graphene-based systems would allow to examine them with the use of more sophisticated tools, analyzing their electronic structure and their subsequent properties.\\
For abovementioned reasons, we decided to consider big graphene-like systems and perform computer simulations using empirical method. We investigate binary $B_xC_{1-x}$, $N_xC_{1-x}$ alloys in vast range of concentrations of B(N) atoms (up to 50\%). Moreover, we extend our research to systems simultaneously doped with boron and nitrogen i.e. $B_{x}C_{1-x-y}N_{y}$ hexagonal structures. Using Tersoff potential (described in \Cref{section:Tersoff_descr}) and Monte Carlo method in Metropolis approach we determine most energetically favorable distribution of B and N within graphene lattice for all cases. Results are shown in  \Cref{subsection:Binary} and \Cref{subsection:Ternary}. To provide qualitative study of investigated systems, in \Cref{subsection:W-C_results} we analyze the results in terms of Warren-Cowley SRO parameters (shortly introduced in \Cref{subsection:Warren-Cowley}). This allows to draw conclusions on order existing in binary and ternary B-C-N alloys. 

\section{Tersoff potential for simulating boron carbon nitrides}
\label{section:Tersoff_descr}

For the studies on ordering in C-B-N alloys we decided to employ Tersoff potential \cite{e[2.1]}. It belongs to the group of bond-order potentials, widely utilized in molecular mechanics and molecular dynamics simulations. Strength of interaction between two atoms there is influenced by their surroundngs, resulting from the form of potential. This leads to more correct description of chemical reactions, compared to simple harmonic approximation e.g. in Keating potential \cite{e[2.2]}.\\
In present work, we use Tersoff potential in the form adjusted for multicomponent systems following \cite{Kroll}, that is a sum over two-center interatomic contributions $V_{ij}$ 
\begin{equation}
\label{eq:Tersoff1}
V_{ij} = f_C (r_{ij}) \left( f_R(r_{ij}) + b_{ij} f_A(r_{ij})\right),
\end{equation}
where $f_R$ and $f_A$ describe repulsive and attractive interactions respectively. $f_C$ is cut-off function, that ensures potential to be short ranged
\begin{equation}
f_A (r_{ij}) = A_{ij} e^{-\lambda r_{ij}},\>\>\>\>\> f_R (r_{ij}) = B_{ij} e^{-\mu_{ij} r_{ij}},
\end{equation}
\begin{equation}
f_C(r_{ij})= \left\{ \begin{array}{l}1 ,\>\>\> r_{ij} < R_{ij}\\ \frac{1}{2} + \frac{1}{2} cos\left( \frac {\pi (r_{ij}- R_{ij})}{(S_{ij} - R_{ij})} \right),\>\>\> R_{ij}< r_{ij} < S_{ij}'. \\0,\>\>\> r_{ij} < S_{ij} \end{array} \right.
\end{equation} 
 The bond-order term $b_{ij}$ in \Cref{eq:Tersoff1} modifies attractive part of potential and describes the influence of surroundings on the binding strength 
\begin{equation}
b_{ij} = \chi_{ij} \left( 1 + \zeta_{ij}^{n_{i}} \right)^{-\frac{1}{2n_i}},
\label{eq:beta_ijk}
\end{equation}
\begin{equation}
\zeta_{ij} = \sum_{k \neq i,j} \>f_C(r_{ik})\> \omega_{ik} \>\beta_i\>g(\theta_{ijk}),
\end{equation}
\begin{equation}
\label{eq:Tersoff2}
g(\theta_{ijk}) = 1 + \frac{c_i^2}{d_i^2} -   \frac{c_i^2}{d_i^2 + \left(h_i - cos(\theta_{ijk})\right)^2} .
\end{equation}
Here, $\theta_{ijk}$ is the angle between bonds $ij$ and $ik$. There occur 13 different parameters in Tersoff potential, each of them dependent on the types of $i,j$ atoms, determining potencial energy of the system in any configuration. Values of the parameters are chosen arbitrarily, on the basis of fitting procedures to describe properly a range of structures. In present work we use parameters fitted for boron carbon nitride systems, following \cite{Matsunaga}.


\section{Theoretical description of order in alloys}

       It is quite easy to intuitively understand the concepts of order and disorder. However, for the scientific purposes we need to provide precise definitions, and then it appears that there is no unique way to understand and describe ideas of ordering. Various theories were developed to give predictions of both short- and long range order, among them: Bragg and Williams\cite{Bragg-Williams}, Bethe\cite{Bethe}, or Peierls\cite{Peierls} theories. Here, we shortly introduce Warren and Cowley\cite{Cowley} approach, that gives description of order in short range, relevant for functionalized graphene systems that are in the limelight.

\subsection{Warren-Cowley Short Range Order parameters}
\label{subsection:Warren-Cowley}

The method proposed by Warren and Cowley is based on family of short range order parameters $\Gamma^{(i)}_{AB}$. They are defined to describe preference of atoms in multi-component system to occur in each other neighbourhood \cite{Cowley}. $\Gamma^{(i)}_{AB}$ is closely related to probability $P^{(i)}_{AB}$ of finding atoms of types A and B as ${i}^{th}$ neighbours compared to the concentration $c_A$ in an alloy
\begin{equation}
\label{eq:gam_def}
\Gamma^{(i)}_{AB} = 1 - \frac{P^{(i)}_{AB}}{c_{A}}
\end{equation}

When calculated, value of $\Gamma^{(i)} _{AB}$  gives us information about preference or antipreference of A and B type atoms to be neighbors in the $i^{th}$ coordination shell:\\

$ \Gamma^{(i)} _{AB} > 0 $         :  atoms of type A tend to not occupy the $i^{th}$ shell of B atoms;

$  \Gamma^{(i)} _{AB} \approx 0 $         :  atoms of type A are neutral to occupy the $ i^{th}$ shell of  B atoms; 
$  \Gamma^{(i)} _{AB} < 0 $        :  atoms of type A tend to ocupy the $ i^{th}$ shell of  B atoms.
\newline\newline
Further analysis reveals symmetry relation $\Gamma_{AB} = \Gamma_{BA}$. Moreover, the following sum rule holds
\begin{equation}
\label{eq:rule2}
\sum_{D} c_D  \Gamma^{(i)} _{CD} = 0,
\end{equation}
for any component C and for any coordination shell\cite{Ducastelle}. Consequently, W-C parameters are not independent. In binary case, there are four parameters $\Gamma_{AA}, \Gamma_{AB}, \Gamma_{BA}, \Gamma_{BB}$, yet due to above rules, there is only one to be independent. In ternary alloys, we obtain three independent parameters. Finally, we may reformulate the conceptual definition from \Cref{eq:gam_def} to the form more convenient for calculations. $\Gamma_{AB}$ can be expressed via number $N_{AB}$ of $A-B$ bonds in the investigated system, related to concentrations $c_A , c_B $ of atoms of type A and B respectively and to the total number of bonds in the system  $N_\text{tot}$ as following
\begin{equation}
\Gamma^{(i)} _{AB} = 1 - \frac{(1 + \delta_{AB})}{2 c_A c_B} \frac{N_{\text{AB}}}{N_\text{tot}}.
\label{equation:WC_bonds}
\end{equation}
\newline
with $\delta_{AB} = 1$ when A and B are different species,  $\delta_{AB} = 0$ otherwise. 

\section{COMPUTATIONAL DETAILS AND RESULTS}

Calculations of binary and ternary alloys have been performed for systems composed of 100 x 100 graphene primitive cells (20 000 atoms), with in-plane periodic boundary conditions assumed. Positions of atoms were fixed during simulation, thus implicitly we restricted ourselves to studies of the chemical order. It was ascertained, that for the chosen size of the system, results do not depend on boundary conditions and that the size is sufficient to reach satisfactory convergence. In order to determine equilibrium configuration of atoms in the studied systems, Monte Carlo (MC) technique in the static, Metropolis NVT ensemble has been applied. Specifically, in single MC step a random pair of atoms with different types is chosen and their positions are switched. Difference in energy $\Delta E $ before and after the swap affects acceptance rate of the change $p = \exp(\frac{ - \Delta E}{k_B T} )$. MC simulations have been performed for wide range of concentrations and temperatures from 70 K up to 1500 K. 
For every combination of those, $M = 10 $ simulations were performed allowing for reasonable  statistics and reducing influence of accidental fluctuations. 

All the simulations are performed using Monte Carlo code designed and developed for the purpose of this research.  The discussion of results is performed in the following section. First, we demonstrate results for the equilibrium configurations and their energy, then we turn to analysis of Warren-Cowley parameters.

\subsection{Binary Alloys}
\label{subsection:Binary}

\begin{figure}[h]
\centering
\caption{Initial positions (left side) and final positions (right) in the system with 2\% (upper row), 10\% (middle) and 40\% (bottom) of boron atoms for simulation with T = 500K. Red dots remark boron, grey denote carbon atoms.}
\includegraphics[width=0.4\textwidth]{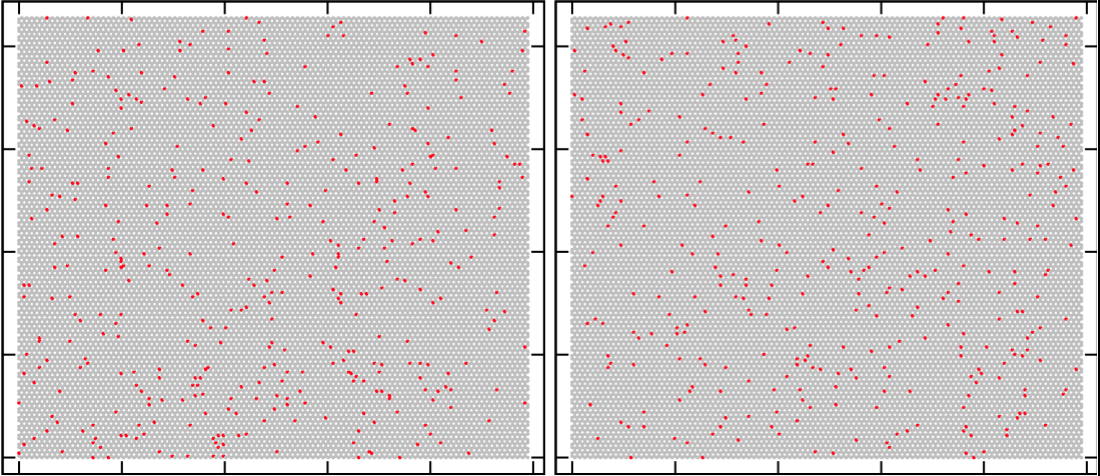}
\includegraphics[width=0.4\textwidth]{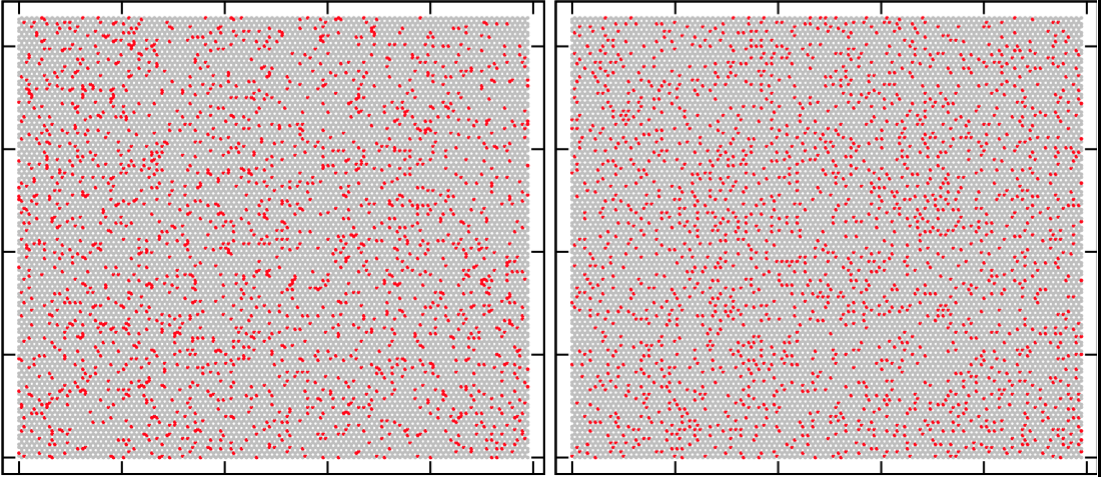}
\includegraphics[width=0.4\textwidth]{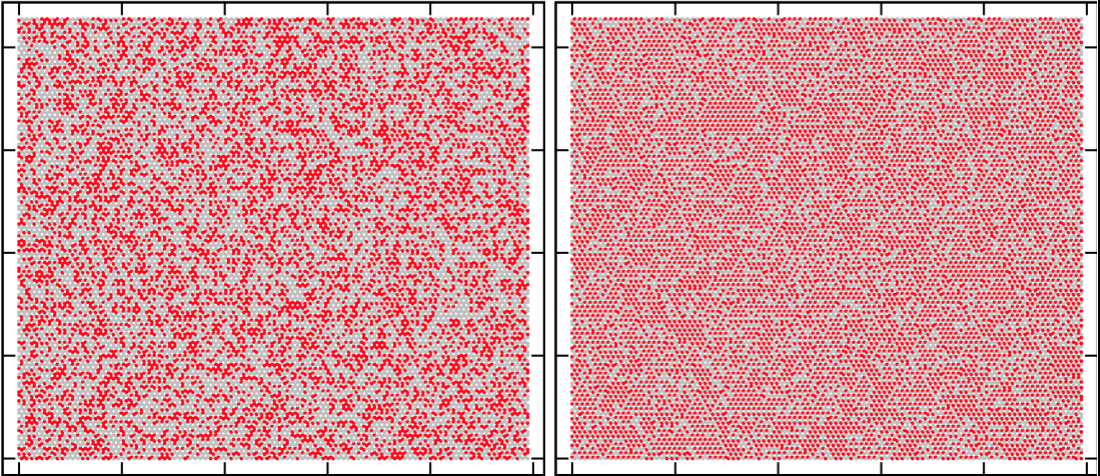}
\label{figure:CB_results1}
\end{figure}

\begin{figure}[h]
\begin{center}
\caption{Initial positions (left side) and final positions (right) in the system with (upper row) 2\%, (middle) 10\% and 40\% (bottom) of nitrogen atoms for simulation with T = 500K. Blue dots remark nitrogen, grey denote carbon atoms.}
\includegraphics[width=0.4\textwidth]{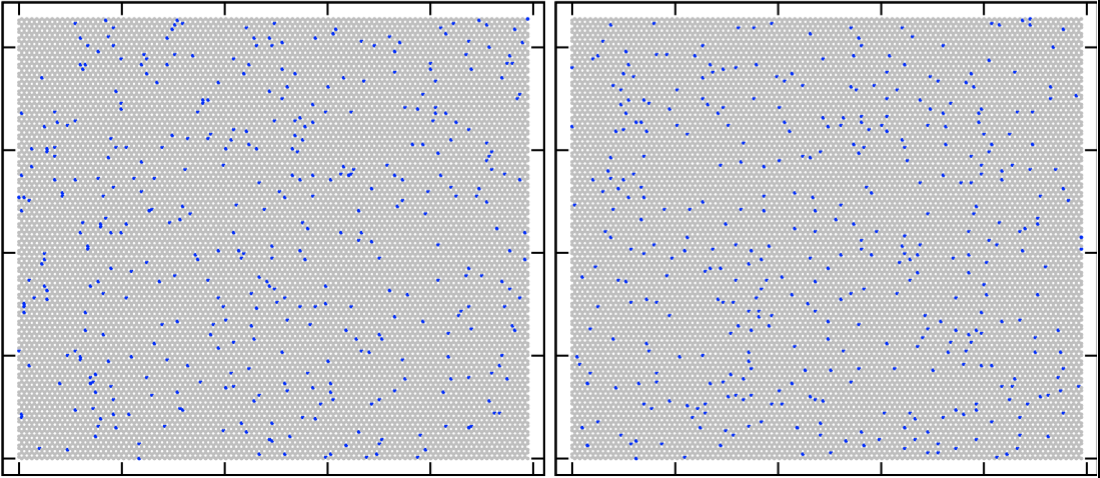}
\includegraphics[width=0.4\textwidth]{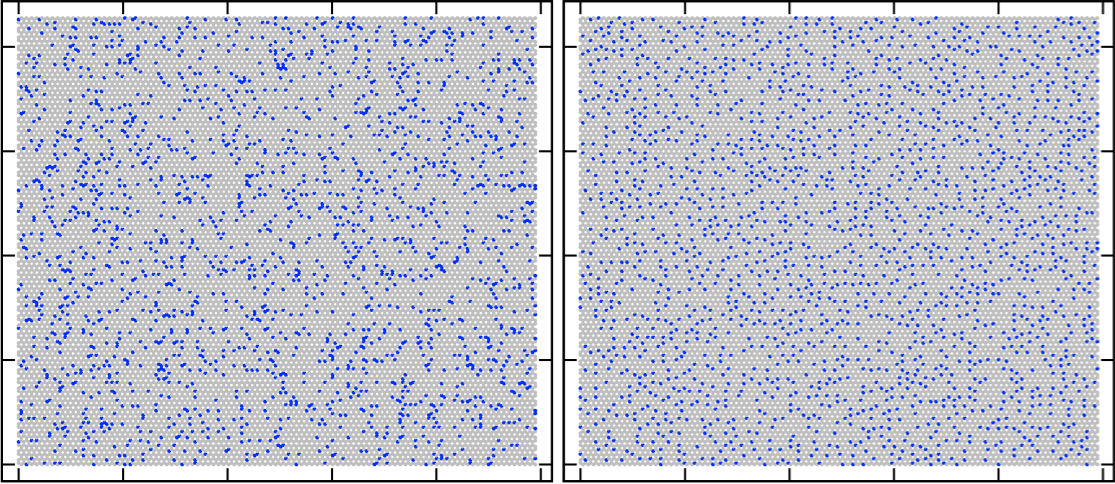}
\includegraphics[width=0.4\textwidth]{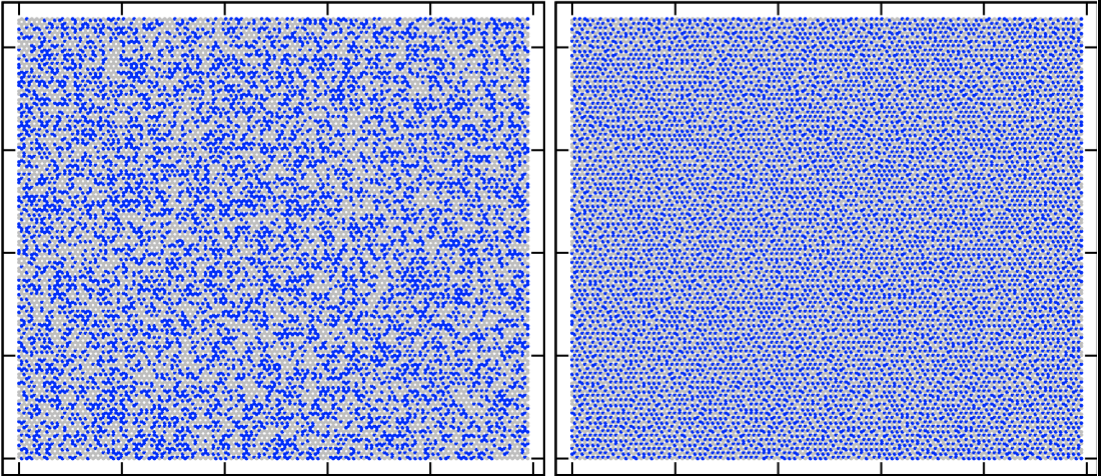}
\label{figure:CN_results1}
\end{center}
\end{figure}

\Cref{table:MC_parameters} summarizes the parameters used during MC simulations of $C_{1-x}N_{x}$ and $C_{1-x}B_{x}$ alloys. For every simulation random initial configuration of atoms was chosen. To provide some intuitions on the nature of investigated systems, we present initial and final distribution of B (N) atoms after $2 \cdot 10^{6}$ MCS for exemplary concentrations in  \Cref{figure:CB_results1} and \Cref{figure:CN_results1}.

\begin{table}[h]
\caption[c]{Parameters of Monte Carlo simulations for binary systems.}
\begin{center}
\begin{tabular}{|l|l|}
\hline
System:                           & 100x100                                               \\ \hline
Number of atoms:            & 20 000                                           \\ \hline
Temperatures {[}K{]}:             & 70,  150,  300,  500,  800,  1000,  1200,  1500                  \phantom{.}   \\ \hline
Substituting atoms:    & Nitrogen (N), Boron (B)                          \\ \hline
Concentrations {[}\%{]}:           & 1,  2,  4,  8,  10,  20,  30,  40,  50      \\ \hline
MCS :        \phantom{.} &   2 000 000                                          \\ \hline
\end{tabular}
\label{table:MC_parameters}
\end{center}
\end{table}
 
\begin{figure}[H]
\centering
\caption[]{Average gains in energy (per atom) for different temperatures in Monte Carlo simulations of B-graphene as a function of B concentration.\\}
\includegraphics[width=0.5\textwidth]{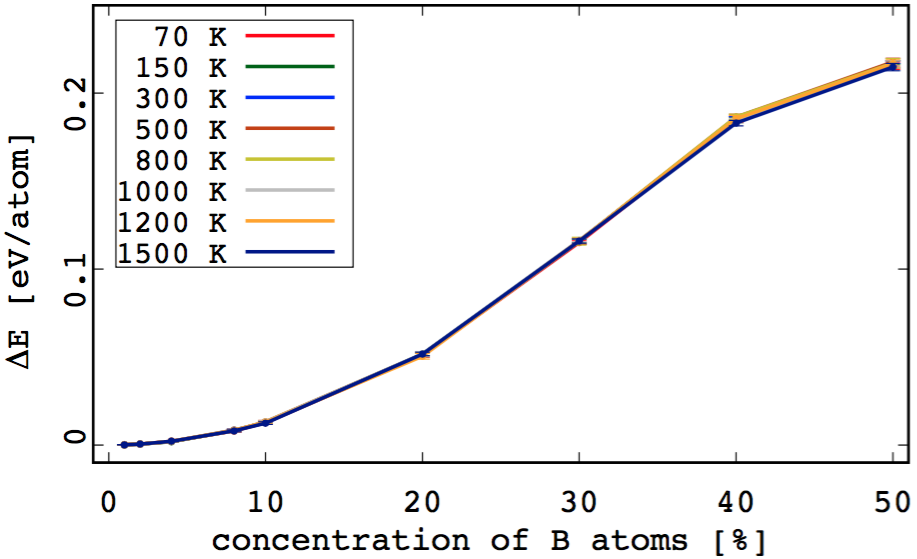}
\label{figure:B_average_deltas} 
\end{figure}
\begin{figure}[H]
\centering
\caption[]{Average gains in energy (per atom) for different temperatures in Monte Carlo simulations of N-graphene as a function of N concentration.\\}
\includegraphics[scale=0.25]{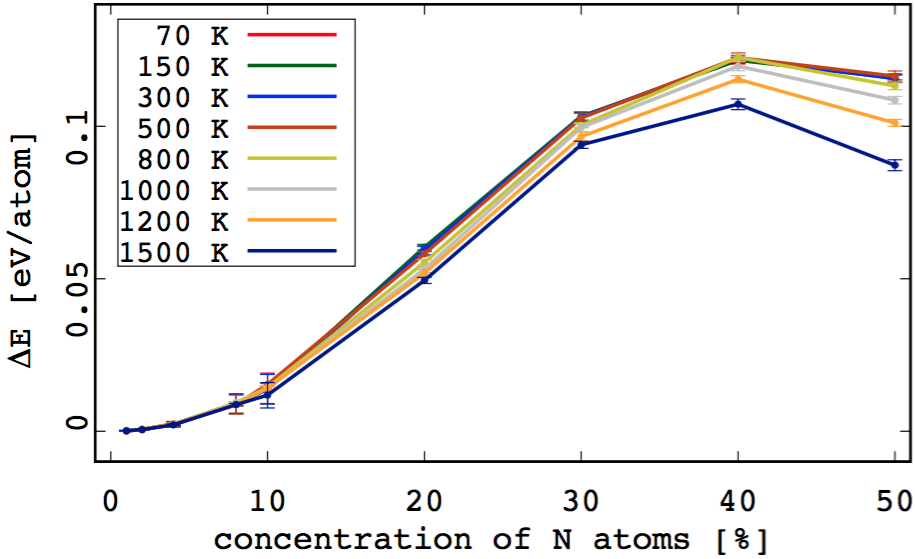}
\label{figure:N_average_deltas} 
\end{figure}

There are only three out of seventy two calculated sets of parameters for each type, however, the others exhibit the same tendency towards uniform distribution of N(B) within the lattice. Energies before and after simulation for all systems were compared and the results are shown in \Cref{figure:B_average_deltas} and \Cref{figure:N_average_deltas}. In B-GL we can observe that with increasing concentration of B, optimization is getting more important. The picture changes slightly for N-type doping. Here for the highest concentration $C_{0.5}N_{0.5}$ energy gain is smaller than in $C_{0.6}N_{0.4}$, which could be caused by the fact that N-N bonds are highly unfavorable but unavoidable in such N-rich systems.

\subsection{Ternary Alloys}
\label{subsection:Ternary}
\begin{figure}[H]
\begin{center}
\caption{Three possible periodic structures of $BC_{2}N$ hexagonal layer, being under investigation: a) ver 1, b) ver 2 , c) ver 3. Yellow dots denote C atoms, pink - N atoms, green - B atoms.}
\includegraphics[scale=0.09]{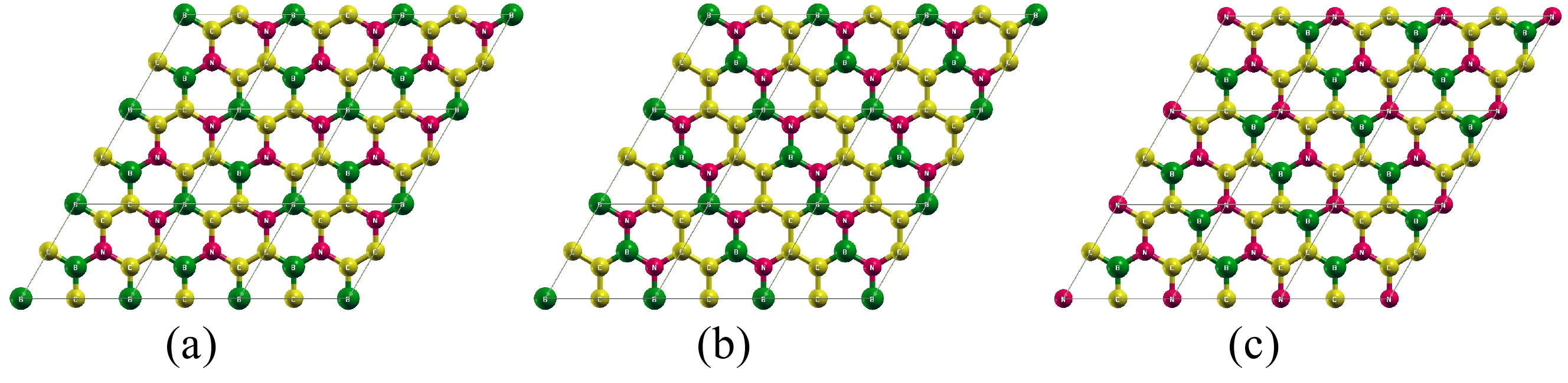}
\label{figure:BC2N_init}
\end{center}
\end{figure}

In case of ternary alloys, we focused on systems containing equal number of boron and nitrogen and we put special atention to systems with 50\% C, 25\% B and 25\% N atoms. Here, in addition to the random initial distribution of atoms, three possible periodic arrangements  (see Ref. \cite{BC2N}) have been investigated. They are shown in \Cref{figure:BC2N_init}. Again for various temperatures, MC simulations were performed. We wanted to examine whether these periodic systems are stable, or realize local (or global) minimum. Summary of parameters used in this part of research is given in \Cref{table:MC_parameters2}.
\begin{table}[h]
\caption[c]{Parameters of Monte Carlo simulations for ternary systems.}
\begin{center}
\begin{tabular}{|l|l|}
\hline
System size:            & 100x100 graphene supercell      \\ \hline
Number of atoms:        & 20 000                          \\ \hline
Temperatures {[}K{]}:   & 70,  150,  300,  500,  800,  1000,  1200,  1500  \phantom{.} \\ \hline
Concentrations:         & 1\%, 5\%, 10\%, 15\% 25\% both N and B        \\ \hline
Initial distribution:   & random (all), ver1, ver2, ver3 (BC$_{2}$N)          \\ \hline
MCS :       			    &   2 000 000                     \\ \hline
\end{tabular}
\label{table:MC_parameters2}
\end{center}
\end{table}

\begin{figure}[H]
\caption{Initial random positions (left side) and final positions (right) in the $B_{0.1}C_{0.8}N_{0.1}$ system for simulation with T = 500K. Red dots remark boron,blue - nitrogen, grey - carbon atoms.}
\centering
\includegraphics[width=0.4\textwidth]{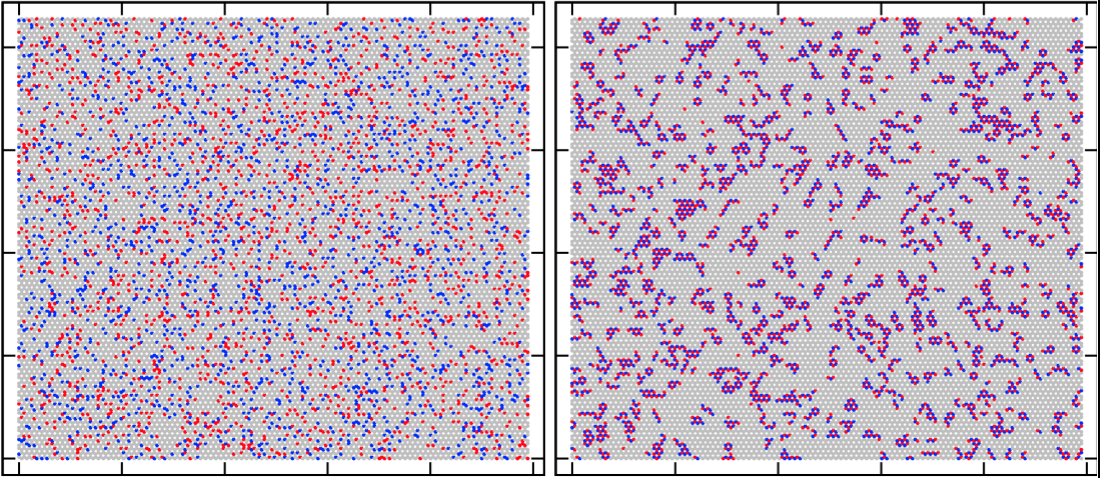}
\label{figure:CBN_results1}
\end{figure}
\begin{figure}[H]
\centering
\caption[]{Average gains in energy (per atom) for different temperatures in Monte Carlo simulations of BN-graphene as a function of B and N concentration for simulations with initial random distribution.\\}
\includegraphics[scale=0.24]{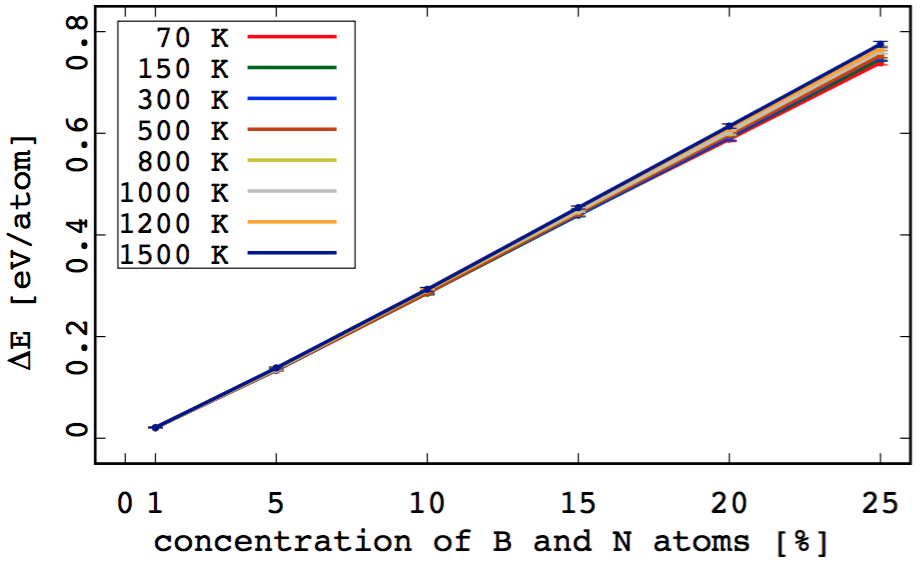}
\label{figure:BCN_average_deltas1} 
\end{figure}
\Cref{figure:CBN_results1} presents initial and final positions of exemplary case of $B_{0.1}C_{0.8}N_{0.1}$ system. As can we see, B and N atoms are very likely to create B-N domains. It is as well the case of all systems with random initial distribution. Here as well, gain of energy averaged over 10 simulations is presented, see \Cref{figure:BCN_average_deltas1}.

\begin{figure}[h]
\begin{center}
\caption{Initial positions (left side) and final positions (right) in the system with random (first row) , ver1 (second), ver2 (third) and ver3 (bottom) initial distribution of atoms. Red dots remark boron,blue - nitrogen, grey - carbon atoms. Simulation performed with T = 500K.}
\includegraphics[width=0.4\textwidth]{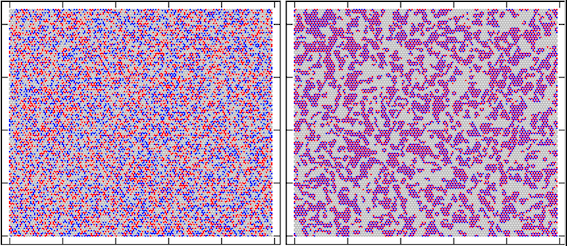}
\includegraphics[width=0.4\textwidth]{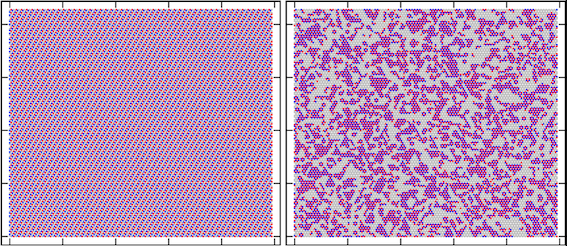}
\includegraphics[width=0.4\textwidth]{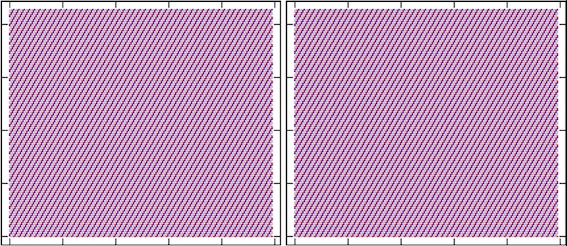}
\includegraphics[width=0.4\textwidth]{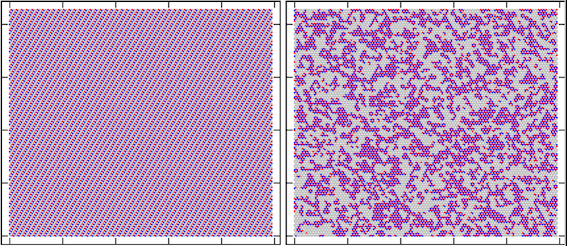}
\label{figure:BC2N_results2}
\end{center}
\end{figure}

\begin{figure}[H]
\centering
\caption[]{Energy per atom during MC simulation for $BC_2N$ systems with different initial distributions for $T = 500K$}
\includegraphics[width=0.5\textwidth]{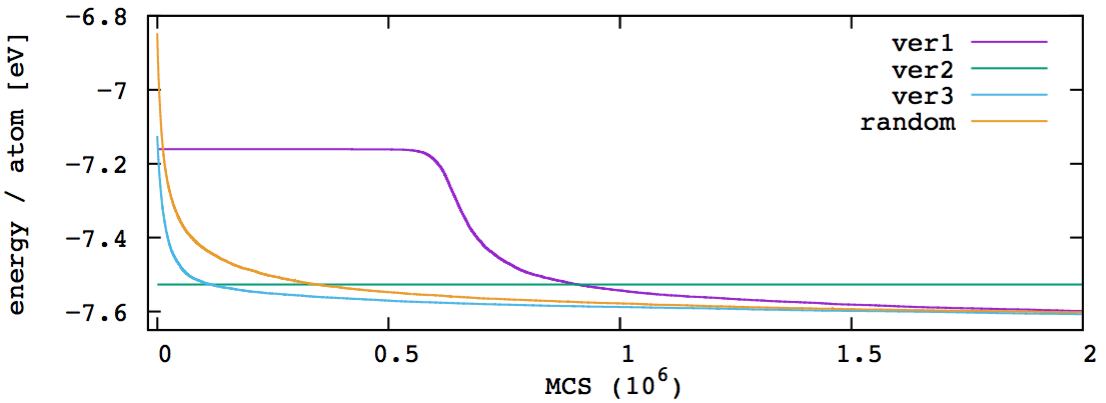}
\label{figure:BC2N_deltas} 
\end{figure}

As was already mentioned, for $B_{0.25}C_{0.5}N_{0.25}$ case, we performed calculations for three initial ordered structures. \Cref{figure:BC2N_results2} shows random and three ordered structures before and after simulation.
Three out of four types convert to a mixture of graphene and B-N domains ($BC_2N$ver1, $BC_2N$ver3 and random) in $T=500K$. However in $BC_2N$ver1 we observe no optimization for temperatures below 300K. It is due to the fact that any pair of atoms that could swap, caused increase of energy in the value that made it almost impossible to accept the change. In higher temperatures exchange probability increases allowing for optimization. $BC_2N$ver2 exhibits barrier against optimization for whole range of temperatures. However, both $BC_2N$ver1 and $BC_2N$ver2 have higher energies than clustered structures. Thus, we can suspect $BC_2N$ver2 to be metastable state in temperatures up to 1500K, and $BC_2N$ver1 up to 300K. For $BC_2N$ver3 no barrier is observed. Energy of systems during simulation for exemplary case in 500K is presented in \Cref{figure:BC2N_deltas}

After presentation of morphology and energies resulting from the calculations, we can go further, to quantitative analysis of order in the systems studied as presented in \Cref{subsection:W-C_results}.

\subsection{Short-Range Order in hexagonal B-C-N aloys}
\label{subsection:W-C_results}

Having equilibrium distributions determined, we turn to investigation of order in B-C-N systems. Average number of bonds each type have been determined (of 10 samples each type) to calculate W-C parameters, using \Cref{equation:WC_bonds}. Only first neighbors are considered here, hence, the superscript indicating the coordination shell is omitted hereafter. 

\begin{figure}[H]
\centering
\caption[]{Average values of Warren-Cowley (a) $\Gamma_{BB}$, (b) $\Gamma_{BC}$, (c) $\Gamma_{CC}$ parameters after Monte Carlo simulations as a function of concentration $c_B$ of boron atoms in graphene lattice, for various temperatures.}
\includegraphics[width=0.5\textwidth]{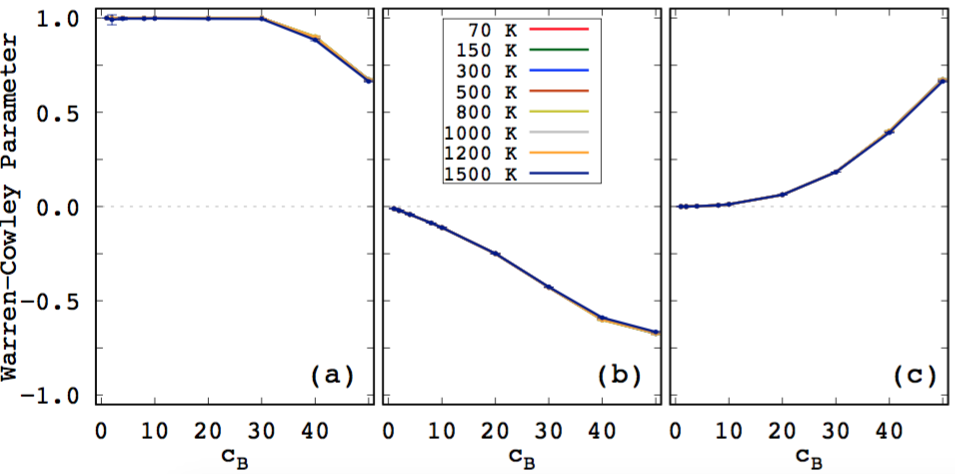}
\label{figure:B_average_W-C}
\end{figure}

\begin{figure}[H]
\centering
\caption[]{Average values of Warren-Cowley (a) $\Gamma_{NN}$, (b) $\Gamma_{NC}$, (c) $\Gamma_{CC} $   parameters after Monte Carlo simulations as a function of concentration $c_N$ of nitrogen atoms in graphene lattice, for various temperatures.}
\includegraphics[width=0.5\textwidth]{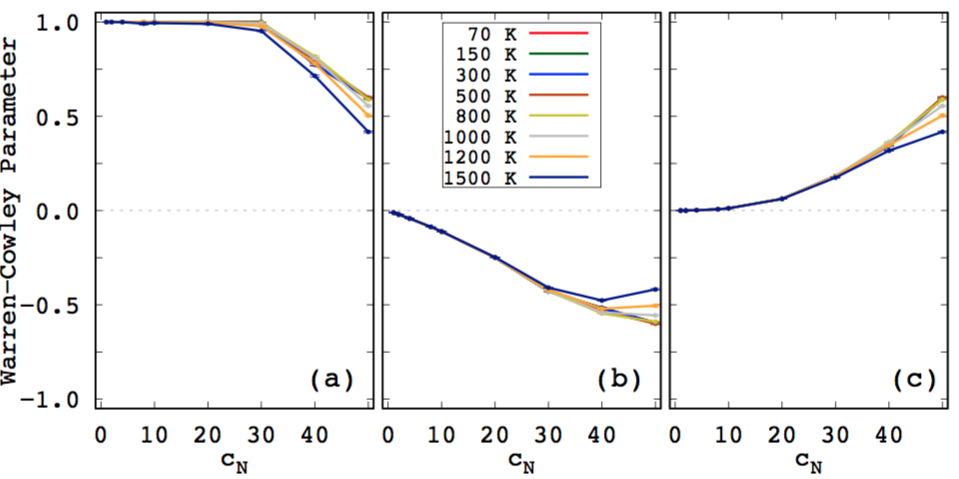}
\label{figure:N_average_W-C} 
\end{figure}

\begin{figure}[H]
\centering
\caption{Average values of Warren-Cowley (a) $\Gamma_{BB}$, (b) $\Gamma_{BC}$, (c) $\Gamma_{CC} $  parameters obtained for initial, random positions of nitrogen atoms for different concentrations $c_B$ in graphene lattice and for different temperatures.}
\includegraphics[width=0.5\textwidth]{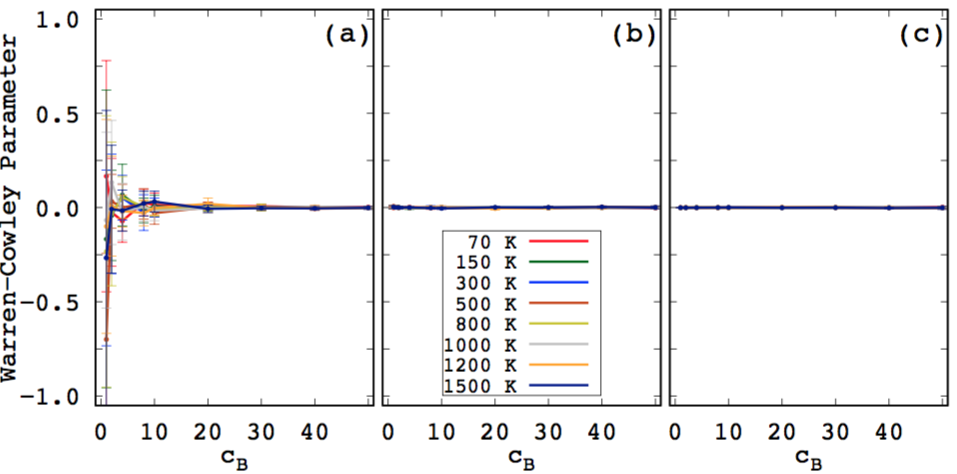}
\label{figure:0_W-C} 
\end{figure}

For binary alloys, we present sets of $\Gamma_{NN}$ ($\Gamma_{BB}$), $\Gamma_{NC}=\Gamma_{CN}$ ($\Gamma_{BC}$) and $\Gamma_{CC}$. However they are interdependent, we present them all in \Crefrange{figure:B_average_W-C}{figure:N_average_W-C} for better picture. We observe both $\Gamma_{NN}$ and $\Gamma_{BB}$ significantly greater than $0$, which means strong tendency of substituting atoms to avoid each other within the lattice. $\Gamma_{BB} = 1$ indicates no $B-B$ bonds in the system and for concentrations up to $30\%$ is really close to that boundary. For highest concentrations it is still much higher than zero. In consequence $\Gamma_{BC} < 0$, proving that $B$ and $C$ "like" each other. Furthermore, we remark no temperature dependence in B-graphene systems. Similar picture arises for N-graphene. Slight variation of the W-C parameters with growing temperature is noticeable there. But even so, for all temperatures we get  $\Gamma_{NN} > 0.4$, which means mutual avoiding of N atoms in hexagonal system. For comparison \Cref{figure:0_W-C} shows parameters for random initial positions for B-GL (N-GL parameters are similar). They fluctuate around zero, which is expected, as probability of finding atoms as neighbors is in accordance with their concentrations.  
We can conclude then that there exist short-range order in binary $N_x C_{1-x}$ and  $B_x C_{1-x}$ hexagonal systems for whole range of investigated concentrations and temperatures. Uniform distribution of substituting 
atoms is energetically favorable and we observe no clustering. 

\begin{figure}[h]
\centering
\caption{Average values of Warren-Cowley (a) $\Gamma_{BN}$, (b) $\Gamma_{BB}$ and (c) $\Gamma_{NN} $ in ternary systems with equal number of B and N atoms after Monte Carlo simulation with random initial positions.}
\includegraphics[width=0.5\textwidth]{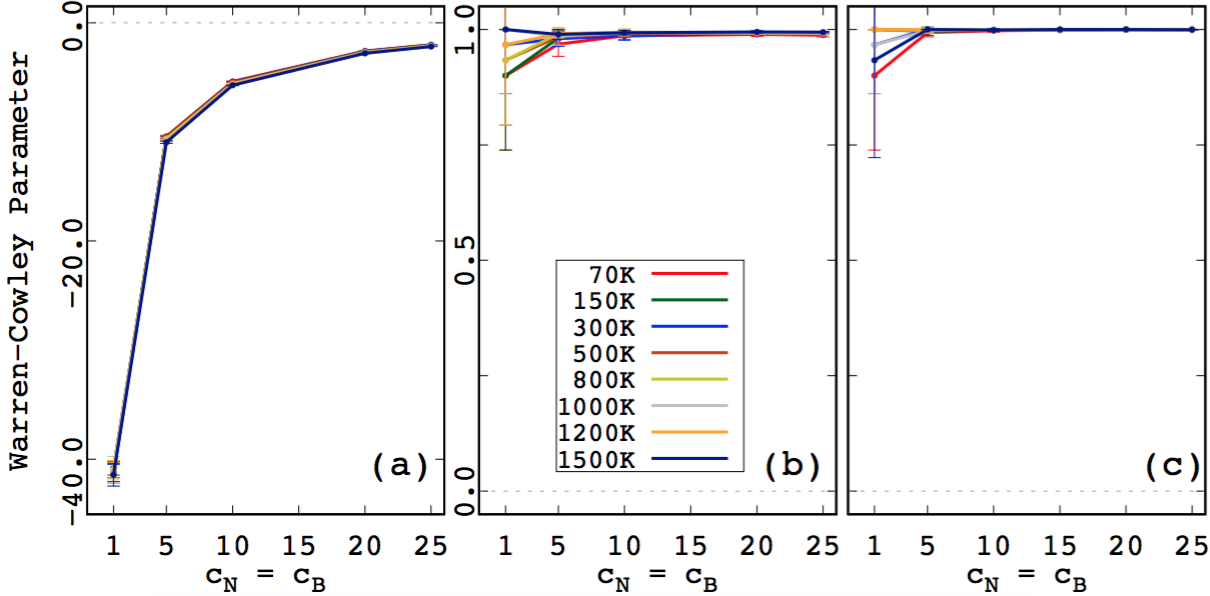}
\label{figure:BCN_W-C} 
\end{figure}

In ternary $B_{x}C_{1-x-y}N_{y}$ alloys there are three independent parameters out of nine (see \Cref{subsection:Warren-Cowley}). We choose  $\Gamma_{BN}$, $\Gamma_{BB}$ and $\Gamma_{NN}$ to wisualize ordering phenomena of investigated systems. Their values are presented in \Cref{figure:BCN_W-C} for a range of concentrations and temperatures. 

As we notice, $\Gamma_{BN}$ is significantly smaller than zero in all the cases: for 1\% each type $\Gamma_{BN} \approx -41$ and for highest concentration around -2. It confirms results form previous part where we could observe B and N forming h-BN clusters. There are also almost no B-B and N-N bonds as in binary case.

\begin{figure}[h]
\centering
\caption{Average values of Warren-Cowley (a) $\Gamma_{BN}$, (b) $\Gamma_{BB}$ and (c) $\Gamma_{NN} $ for various initial distribution in $BC_2N$ hexagonal strucutres. Dark triangles denote initial values of parameters, lines correspond to simulation results performed in various temperatures.}
\includegraphics[width=0.5\textwidth]{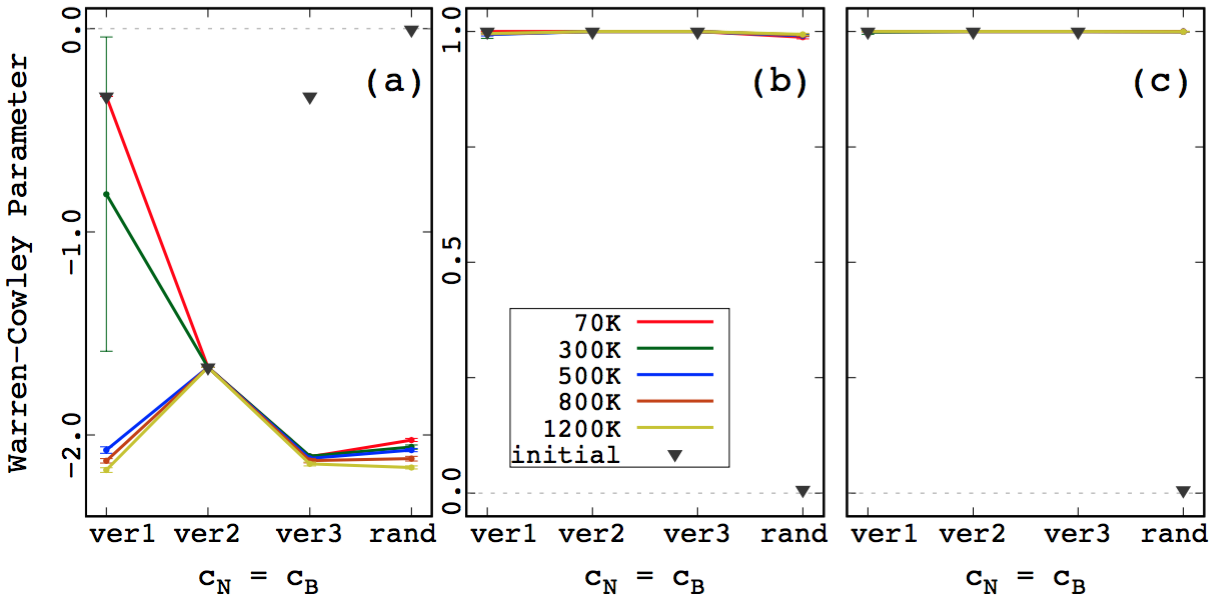}
\label{figure: BCNvers_WC} 
\end{figure}

Finally, we focus on $BC_2N$ case and compare parameters for systems with various initial distributions They are presented in \Cref{figure: BCNvers_WC}. Random distribution gives W-C parameters fluctuating around zero, similarly to the case of binary alloys. Three remaining initial distributions are periodic, hence, ordered in terms of Warren-Cowley approach.

As was discussed in previous chapter, $BC_2N$ver2 remains intact during MC simulation, thus corresponding parameters do not change. In case of $BC_2N$ver1 system simulation perform in lowest temperatures do not lead to full optimization, which is visible in \Cref{figure: BCNvers_WC} (a). However, energies of those structures are higher than in clustered BN-graphene. Creating pure carbon and h-BN domains leads to the lowest energy and is also reflected in the values of W-C parameters ($\Gamma_{BN} < 0$). Thus, we can conclude that energetically favorable in hexagonal systems containig boron, carbon and nitrogen is creation of h-BN clusters surrounded by pristine graphene domains.

\section{Conclusions}

We have studied ordering phenomena in binary $C_{1-x}B_{x}$, $C_{1-x}N_{x}$ and ternary $B_{x}C_{1-x-y}N_{y}$ alloys creating two-dimensional, hexagonal structures. We employed empirical Tersoff potential and Monte Carlo simulation technique in order to determine configurations having the lowest energy in temperatures ranging 70-1500K. We determined Warren-Cowley Short-Range Order parameters to quantify order in those systems In binary case, we investigated structures with B(N) concentration ranging from 1\% to 50\% with random initial positions. After simulation all the systems exhibit tendency to uniform distribution of substituting atoms, which is reflected in values of $\Gamma_{BB}$ $\Gamma_{NN}$ close to one. In ternary case, we investigated $B_{x}C_{1-x-y}N_{y}$ structures with equal number of B and N atoms. Again, after simulations with random initial distribution W-C parameters were determined, confirming that creation of h-BN clusters and surrounded by areas of pristine graphene gives the lowest energy. In special case of $BC_2N$ i.e. 25\% B, 25\% N and 50\% C we examined also three theoretical ordered structures to verify their stability. All of them appear to have higher energies than clustered h-BN-graphene, but two of them may be metastable states - one in low temperatures up to 300K, the other at last to 1500K. To conclude, there exist ordering phenomena in both binary and ternary B-C-N graphene-like systems, and this may have significant impact in further properties of these materials.

\begin{acknowledgments} 
This work has been supported by the NCN grant  HARMONIA  (no.  UMO-2013/10/M/ST3/00793).
\end{acknowledgments}


\bibliography{sample-paper}
\bibliographystyle{prsty}


\end{document}